# First-Principles Analysis on Structural Transition in $BaNi_2As_2$ Superconductor

Wasim Raja Mondal,[1] and Swapan K Pati [1]

Theoretical Sciences Unit[1]

Jawaharlal Nehru Centre for Advanced Scientific Research, Bangalore, 560064, India.

email addresse:

Wasim Raja Mondal: wasimr.mondal@gmail.com

We have studied structural phase transition in $BaNi_2As_2$ superconductor with the application of Phosphorous doping based on first principles calculations. Our results reproduce experimental findings and explain some of the experimental observations.

## Introduction

The discovery of superconductivity with transition temperature ($T_C$) 56K in $RFeAsO_{1-x}F_x$ (R=Sm, Nd, Pr…) [1-6] has attracted attention of many researchers to understand the superconducting mechanism. Besides, the appearance of superconductivity in Ni-based compounds which posses low $T_C$, i.e., LaONiP ($T_C$=3K)[7], LaONiAs($T_C$=2.75 K)[8], $BaNi_2P_2$ ($T_C$=2.4 K)[9], $BaNi_2As_2$ ($T_C$=0.7 K)[10] and $SrNi_2As_2$ ($T_C$=0.62 K)[11] give a new platform to understand the microscopic reasons behind such phenomenon. Hence it is important to understand why $T_C$ is low in Ni-based compounds, which may enrich our understanding about the high $T_C$ in iron-based superconductors. The high $T_C$ in superconductors is sometime associated with structural instability. It has been reported that there is an increase in $T_C$ at structural phase boundary in various systems including cuprates [12, 13], iron-pnictide superconductors [14, 15], A15 compounds [16], graphite intercalated compounds and elements including Li and Te under high pressure [17-19].

Among all the Ni-based pnictides, $BaNi_2As_2$ is mainly important because it is regarded as a non-magnetic analogue of iron-based superconductors e.g $CaFe_2As_2$. Besides these similarities, there is some dissimilarity too. It is reported that collinear spin-density-wave magnetic ordering does not exist in $BaNi_2As_2$ [20]. The absence of magnetism may be one of the reasons for low $T_C$ in $BaNi_2As_2$ [21-24]. This clearly indicates the importance of magnetism in iron-pnictide superconductors for its high $T_C$. $BaNi_2As_2$ has the following properties: (1) It is observed that structural phase transition is first-order in $BaNi_2As_2$ [25]. (2) It is predicted that band-shift in $BaNi_2As_2$ is due to considerable lattice distortion. (3) Correlation effect in BaNi2As2 is very weak since small band renormalization factor in $BaNi_2As_2$. Very recently, it is reported that the structural phase transition in $BaNi_2As_2$ can be tuned by phosphorus doping [33]. Pressure-induced structural phase transition in this system is particularly interesting since it can provide a new way for finding the reason behind the mechanism of high-temperature superconductivity. Based on this interesting report, in this work, we present a first-principle study of $BaNi_2As_2$. Our approach reproduces and explains the experimentally observed changes in $T_C$ with pressure and establishes the relation between the sudden jump in $T_C$ and the structural phase transition. In addition, we address the following questions: (a) how does lattice parameter depend on doping concentration? (b) What is the reason behind doping induced lattice distortion? (c) Is increase in the superconductivity in the tetragonal phase electronic or is it phonon mediated?

## Computational details

Correlation effect in $BaNi_2As_2$ is very low because of small band renormalization factor [20]. Hence non-interacting density functional theory is appropriate in this context. All calculations are performed by considering pseudo potential-based density functional theory (DFT) as implemented in the PWSCF package [61]. We consider generalized gradient approximation (GGA) [62] for the exchange correlation function and use ultra-soft pseudo potential for taking into account the interaction between the ion's core and the valance electrons. We take $3\times3\times1$ super cell for the simulation of $BaNi_2 (As_{1-x}P_x)_2$ with x=0.0, 0.027, 0.055, 0.083, 0.11. Optimization is done with respect to both atomic position and lattice-parameters using Broyden-Fletcher-Goldfarb-shanon (BFGS) method. Plane-wave basis is used with a kinetic energy cutoff of 40 Ry for wave functions and cutoff of 400 Ry for charge density. For the integration over the Brillouin Zone of $3\times3\times1$ super cell containing 90 atoms, we take $3\times3\times3$ Monkhorst-Pack mesh [63]. Convergence is checked carefully in wave-vector space and considered by taking the energy difference between two consecutive steps, $10^{-8}$ Ry. Spin polarized calculation is neglected in all calculation since there is no existence of spin ordering in $BaNi_2As_2$.

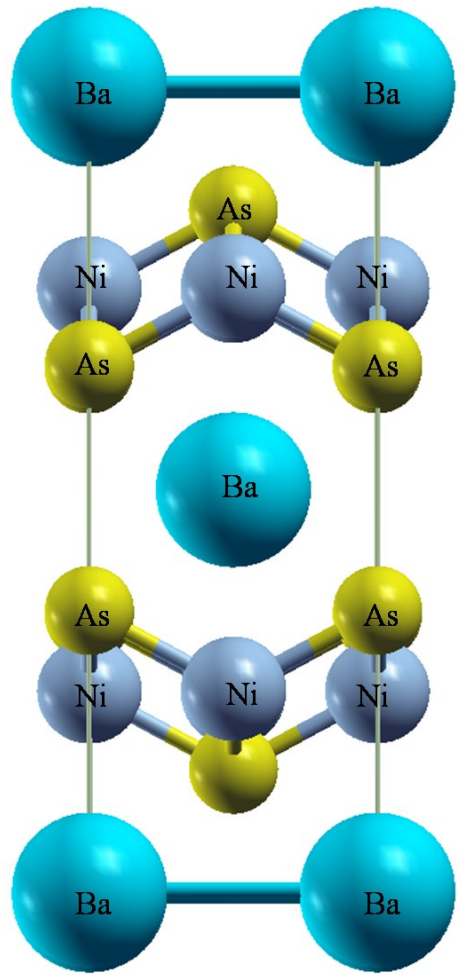

**Figure 1 Crystal structure of BaNi$_2$As$_2$**

**Table 1 Data for structural parameter and atomic position for the unit cell of BaNi$_2$As$_2$ at 297 K**

| Reference | Lattice parameter a(Å) | Lattice parameter c(Å) | $Z_{As}$ |
|---|---|---|---|
| this work | 4.192 | 11.231 | 0.349 |
| Earlier work[34] | - | - | 0.351 |
| Expt.[33] | 4.147 | 11.619 | 0.347 |

# Results and discussion

The unit cell of $Bi_2Ni_2As_2$ at T=293K has the tetragonal structure with I4/mmm symmetry. To verify our structure of $Bi_2Ni_2As_2$, we do variable cell optimization of the unit cell and compare the unit cell parameters of $BaNi_2As_2$ with earlier findings and experimental structure as shown in Table1. Our calculated lattice parameters 'a' and 'c' of the unit cell are 1% larger and 3.3% smaller than experimental values [33] respectively. Our computed $Z_{As}$ (height of the As atom from the bottom of the unit cell) is in good agreement with earlier work [34] with 0.5% smaller value but 0.5% larger than the experimental one [33]. The calculated lattice constants and atomic positions of the unit cell are used as initial values in the super-cell calculations. To explore the pressure dependence on the structure, we do the calculation with various doping concentrations (0.0, 0.027, 0.055, 0.083, 0.11) as described in experimental study [33].

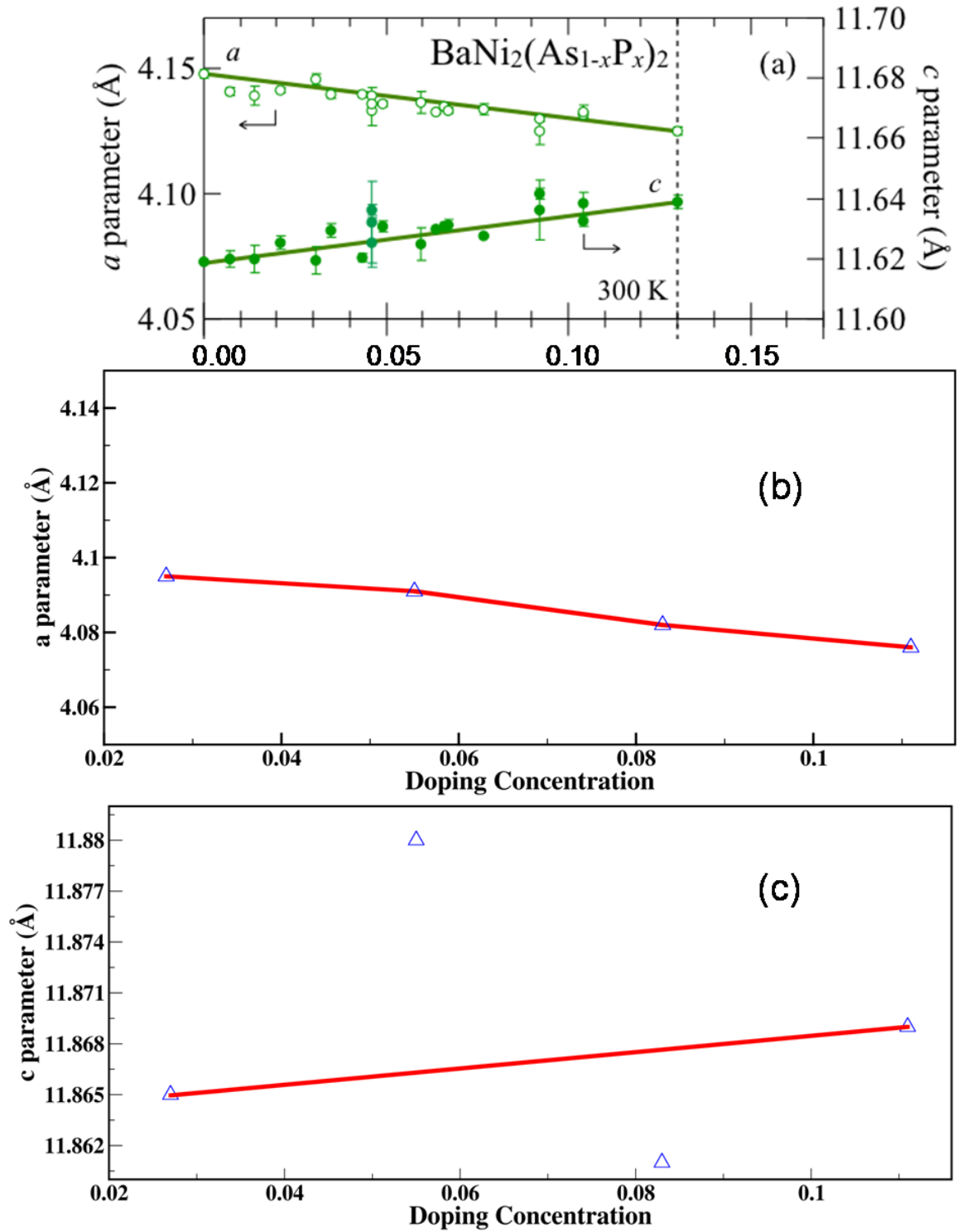

**Figure 2 (a)** Lattice parameters a and c as a function of phosphorous content x for BaNi$_2$(As$_{1-x}$P$_x$)$_2$ as predicted by experiment [33] **(b)** calculated dependence of latiice parameter a with various doping concentration **(c)** calculated dependence of lattice parameter c with various doping concentration

The dependence of lattice parameters 'a' and 'c' with various doping concentration is plotted in figure 2. We observe that the lattice parameter 'a' decreases with increase in doping concentration but lattice parameter 'c' increases with increase in doping concentration. This correlation between lattice parameters and doping concentration is in good agreement with experimental one as shown in figure 2 (a). We also predict the decrease of volume with doping as proposed by the experiment [33]. According to our calculations, decrease in volume is 1.7 $\text{Å}^3$ with 11% phosphorous doping whereas experiments predicted decrease in volume as 1.1$\text{Å}^3$ with 13% phosphorous doping. Given the theoretical approximations, we believe our theoretical results are in reasonable agreement with experimental predictions. The origin of pressure dependence of the volume is as follows: the valance electrons of phosphorous and arsenic are same. But the ionic radius of phosphorous is smaller than arsenic. When we substitute arsenic by phosphorous, this doping creates chemical pressure. That is why; we find this kind of monotonic dependence of lattice parameters with doping concentration and overall volume of the cell. As shown in figure 3(a), with the increase in doping concentrations, energy increases. In other words, doping creates the instability in the structure and the structure tends to go to other phase for stability. From figure 3(b), it is clear that with the increase in doping concentration, the pressure increases. We see structural distortion due to this pressure which appears due to the application of doping. From figure 3 (b), it is interesting to observe that within doping concentration between 0.02 and 0.05 pressure almost increases linearly and in between 0.07 and 0.11 pressure decreases linearly. But in the region between 0.06 and 0.07, pressure remains almost constant. That means, up to doping concentration 0.05, structure becomes unstable in one phase in which it exists. Beyond the range of 0.05, it starts going to another phase. After doping concentration of 0.065, it reaches in one new phase in which it is again stabilizes as pressure decreases.

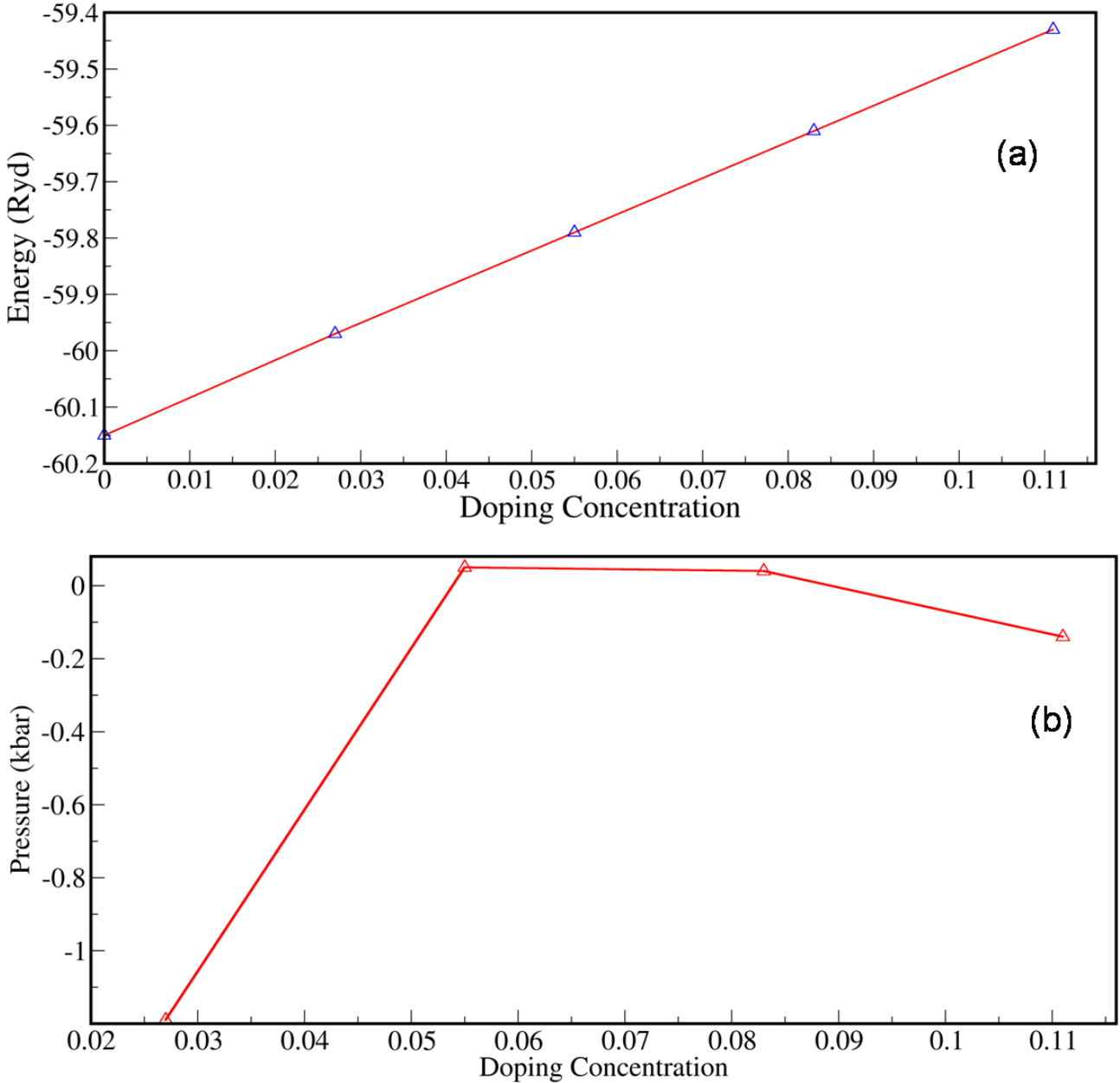

**Figure 3 (a) Energy variations with various doping concentrations (b) variation of pressure with various doping concentrations**

In between 0.06 and 0.07 doping concentrations, pressure remains almost independent of doping concentration. So we can think of structural phase transition occuring in between 0.06 and 0.07 doping concentrations. In fact, experimentally it is observed that the structural phase transition occurs at the doping concentration, 0.067. With the application of doping, besides the lattice parameters, internal parameters also change. This is shown in figure 4. In figure 4, it is clearly seen the bond angle, < Ni-As-Ni, changes with the increase in doping concentration. But it is again interesting to note that there is sudden jump in the change of the bond angle in between doping concentration 0.06 and 0.07. This suggests that the structural phase transition occurs somewhere between doping concentration 0.06 and 0.07 which is in good agreement with experimental findings [33].

**Table 2 Data for density of states $N(E_F)$ at the Fermi level with various doping concentrations**

| Reference | Doping concentration | $N(E_F)$ $(eV)^{-1}$ |
|---|---|---|
| This work | 0.00 | 7.18 |
| Expt.[33] | 0.00 | 6.00 |
| Singh et al.[34] | 0.00 | 7.14 |
| This work | 0.25 | 7.19 |
| Expt.[33] | 0.25 | 6.00 |

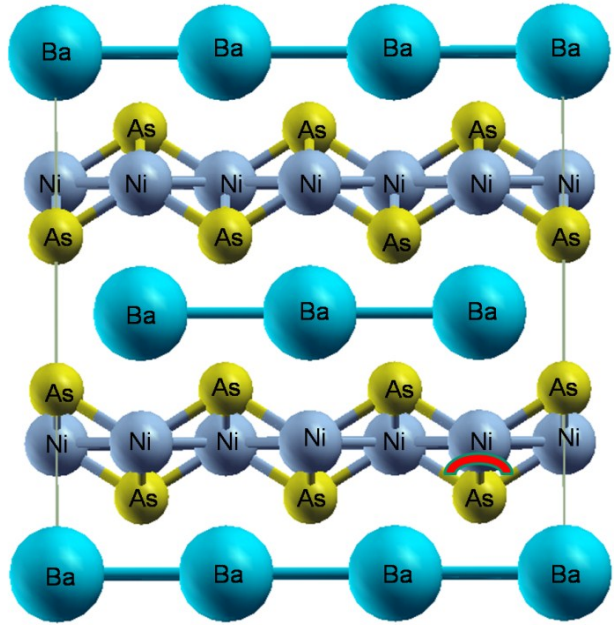

**Figure 4 3× 3×1 super cell of BaNi$_2$(As$_{1-x}$P$_x$)$_2$**

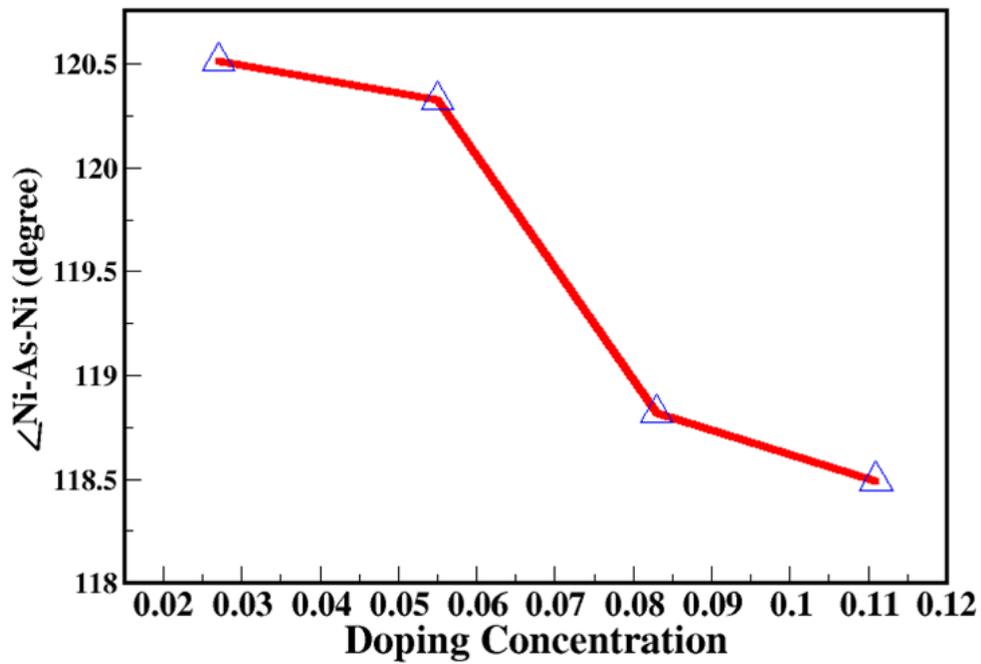

**Figure 5 Change in the bond angle with doping concentration**

Thus, there is structural phase transition in BaNi$_2$(As$_{1-x}$P$_x$)$_2$ due to the application of doping. With this structural phase transition, there is increase in T$_C$ which is the main advantage of this transition. Now the question is what is the origin of this enhancement in the value of T$_c$? Is it related with electron or phonon? To find out the reason, we have plotted density of states (DOS) in various doping concentrations which is shown in Fig. 6. There is finite density of states at the Fermi level in both doping concentration (x=0.0 and x=0.25). So the parent composition, BaNi2(As$_{1-x}$P$_x$)$_2$, superconductor is metallic like iron-pnictide superconductors and this metallic behavior of BaNi$_2$(As$_{1-x}$P$_x$)$_2$ is independent of the amount of doping. From the DOS at x=0.0 doping concentration, we have computed the density of state at the Fermi level, N (E$_F$), which is in good agreement with earlier work and experimental prediction. All these values are tabulated in Table 2. We have also calculated N (E$_F$) for the higher doping concentration. This N (E$_F$) value for the higher doping concentration (x=0.25) is also in agreemant with experimental value. Interestingly, calculated values of N(E$_F$) are independent of the doping concentrations.

$$\gamma = \frac{\pi^2}{3} N(E_F) k_B^2 \qquad (3.1)$$

Where $\gamma$ is the electronic specific heat coefficient, $k_B$ is the Boltzmann constant and N(E$_F$) is the density of state at the Fermi level.

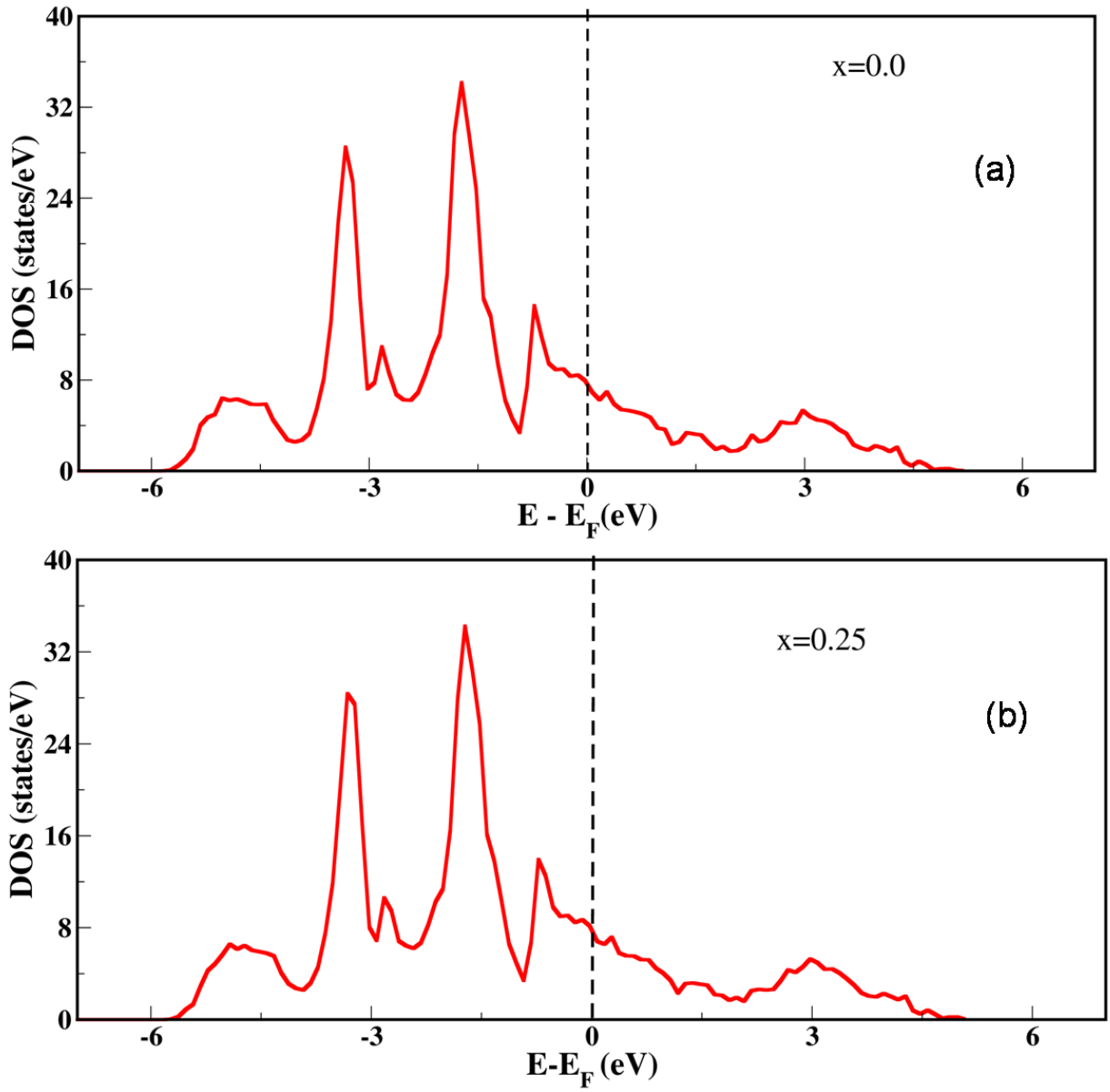

**Figure 6** Density of state plot for various doping concentrations (a) 0.00 (b) 0.25

$$\gamma^{expt} = \gamma^{theor}(1 + \lambda_{ep}) \qquad (3.2)$$

Where $\lambda_{ep}$ is the electron-phonon coupling constant.

$$k_B T_c = \frac{\hbar \omega_{ln}}{1.2} \exp\left\{ -\frac{1.04(1+\lambda_{ep})}{\lambda_{ep} - \mu^*(1+0.62\lambda_{ep})} \right\} \qquad (3.3)$$

Where $\omega_{ln}$ is the logarithmic average of phonon frequency and $\mu^*$ is a constant and equal to 0.12

From equation (3.1), it is clear that electronic specific- heat coefficient ($\gamma$) is directly related with $N(E_F)$. Basically, it is only a function of $N(E_F)$ within free electron theory. Now, if $N(E_F)$ is independent of the amount of doping, the value should be remain constant with doping. This explains the experimental prediction of the variation of specific heat with temperature. According to experimental observations, the slope of the C/T vs. $T^2$ plot is almost unchanged with various doping concentration. This indicates the independent value of electronic specific-heat coefficient ($\gamma$) with various doping concentrations. From equation (3.2), electron-phonon coupling constant is roughly related with $\gamma$. If $\gamma$ is constant against doping concentration, then electron-phonon coupling is also constant with various doping concentration. From Allen-Dynes formula, as given in equation (3.3), it is clear that transition temperature is a function of logarithmic average of phonon frequency $\omega_{ln}$ and electron-phonon coupling constant $\lambda_{ep}$. Now if $\lambda_{ep}$ remains almost constant with various doping concentrations, then the only possibility for the increase in transition temperature ($T_C$) is $\omega_{ln}$. Hence, the enhancement in $T_C$ is related with phonons, not with electrons. Our finding for the origin of the enhancement in the superconductivity is in complete agreement with the experimental predictions.

## Conclusion

We have presented a systematic first-principles study on the structural phase transition in BaNi$_2$(As$_{1-x}$P$_x$)$_2$ superconductor. Our calculated dependence of lattice parameters with various doping concentrations is in well agreement with experimental findings. We have also found overall decrease in volume with increase in the doping concentrations, which compares fairly well with experimental findings. We have seen distortion in the internal parameters like bond angle (<Ni-As-N) changes significantly with increase in the doping concentrations. We have clearly found the reasons behind the crystal distortions with doping. We have shown phosphorous doping creates pressure on BaNi$_2$(As$_{1-x}$P$_x$)$_2$ and how this induced pressure produces instability in the system. We have predicted the enhancement in the increase of transition temperature (T$_C$) is related with phonons not with electrons.